\documentclass[12pt]{article}
\usepackage{amssymb,epsfig}
\begin{document}

\title{
Spectroscopic properties 
of a two-level atom interacting with a complex
spherical nanoshell
}

\author{
Alexander Moroz\thanks{http://www.wave-scattering.com}
\\
Wave-scattering.com
} 

\maketitle


\begin{center}
{\large\sc abstract}
\end{center}
Frequency shifts, radiative decay rates, 
the Ohmic loss contribution to the nonradiative decay rates,
fluorescence yield,
and photobleaching of 
a two-level atom radiating anywhere inside or outside 
a complex spherical nanoshell, i.e. a stratified sphere 
consisting of alternating silica and gold concentric spherical shells, 
are studied.
The changes in the spectroscopic properties of an atom interacting with
complex nanoshells are significantly enhanced, often 
more than two orders of magnitude, compared
to the same atom interacting with a homogeneous 
dielectric sphere. The detected fluorescence intensity 
can be enhanced by $5$ or more
orders of magnitude. The changes strongly depend
on the nanoshell parameters and the atom position. 
When an atom approaches a metal shell,
decay rates are strongly enhanced 
yet fluorescence yield exhibits a well-known
quenching.
Rather contra-intuitively, the Ohmic loss contribution 
to the nonradiative decay rates
for an atomic dipole within the silica core 
of larger nanoshells may be decreasing when
the silica core - inner gold shell interface is approached.
The quasistatic result that 
the radial frequency shift in a close proximity
of a spherical shell interface is approximately 
twice as large as the tangential frequency shift appears to 
apply also for complex nanoshells.
Significantly modified spectroscopic properties 
(see computer program freely available at http://www.wave-scattering.com)
can be observed in a broad band
comprising all (nonresonant)  optical
and near-infrared wavelengths.

\vspace*{0.6cm}

PACS numbers:
78.67.Bf, 33.70.Jg, 32.70.Jz, 33.50.-j, 87.64.Ni, 87.64.Xx \hfill

\vspace*{1.9cm}

\newpage

\section{Introduction}
Spectroscopic properties 
of an isolated atom, such as radiative and nonradiative
decay rates, frequency shifts, and fluorescence yields
are not inherent to the atom but characteristic of an atom 
coupled to a physical system.
Indeed, it has been known for a long time that the presence 
of a small structure, cavity, or an interface 
can significantly impact the characteristic behaviour of a
radiating system, irrespective if the
emission takes place inside or adjacent to a material body \cite{Pu,RiK,LRG}. 
The physical origin of the decay of an excited atom state
is the coupling of the atom 
to the vacuum electromagnetic field. A nearby presence of a
material body modifies 
the vacuum electromagnetic field at the atom position.
Consequently, the atom interacts with the modified 
vacuum electromagnetic field and it will exhibit different spectroscopic
properties than the same atom in the absence of the material body.
There is growing interest in the application of various
systems that can significantly affect the
vacuum electromagnetic modes. 
Such systems are currently of great interest in 
the fields of photonics and quantum electrodynamics. 
They have found widespread application
in microcavity lasers, electroluminescent
devices, and proposed photonic band-gap devices.
There is also growing interest in the application of various
aspects of the molecule–surface interaction in the field of
medical diagnostics, particularly in the immuno-assay
area, in which fluorescence-based techniques are widely
used \cite{Lak}. Alternatively, for the applications in near-field 
optical microscopy, one is interested 
in changes in the atom 
fluorescence properties induced by the presence of a nearby 
microscope tip.

For the purpose of this article, the atom would mean any localized
fluorescent dipole source, e.g., 
fluorescent organic group, rare earth atom, etc. 
The  atom would be considered as a 
two-level system in the regime of weak
coupling, within the domain of applicability of the 
linear response theory \cite{AgaIV,WyS,DKW,DKW1}. 
In the latter case, the quantum-mechanical 
description \cite{AgaIV,WyS,DKW,DKW1} yields identical results
to the classical description \cite{CPS,Ch,Chew,AMap,KDLm,KDL1}.
The spectroscopic properties 
of the  atom will be studied as a function of the atom position
inside and outside stratified spheres.
The theory of fluorescence properties of an atomic dipole has
has been mostly investigated only in the case of a homogeneous sphere 
\cite{DKW,DKW1,Ch,Chew,KDLm,KDL1,Rup}.
Chew et al \cite{CKN} provided a formal solution to the
problem of a dipole radiating in the presence of a multilayered 
sphere. However, their solution for the sphere with 
$N$ concentric shells (the sphere
core counts as shell number one) is written in terms of a 
$2N\times 2N$ matrix
and appears awkward and impractical for numerical calculations.
Indeed, neither Chew nor anybody else
have appeared to implement the Chew et al \cite{CKN} solution numerically. 
The main obstacles are that
as $N$ increases so do computer memory requirements to store the
matrix and the time to carry out the matrix calculations, which 
increases as $N^3$. 
Whenever a radiating dipole has been discussed interacting
with a multicoated sphere, either the problem has been treated
in a quasi-static approximation \cite{LRG}, or the dipole has only
been allow to
radiate at the sphere origin \cite{DKW,SuHII,Tom,KLa,End}.
The dipole position in the center of such a complex sphere
considerably simplifies calculation as the inherently
vector problem reduces to the scalar one involving  scalar fields
$ {\bf r} \cdot{\bf B}({\bf r})$ and 
${\bf r}\cdot{\bf D}({\bf r})$ \cite{SuHII,Tom}.
Additionally, in most cases only the simplest core-shell particles
have been dealt with \cite{LRG,DKW,KLa,End}. 
Although Li et al \cite{LKL} has provided
a recursive formula for Green's function for the case
of a multicoated sphere with an arbitrary number of concentric shells, 
an efficient numerical use of Li's formula, even 
for coinciding spatial arguments, 
requires to perform
traces over the magnetic angular momentum number.
Only very recently the limitation on the dipole position at
the center of a general multilayered sphere has been removed 
and  the traces over the magnetic angular momentum number
in the scattering Green's function at coinciding
spatial arguments, ${\bf G}({\bf r},{\bf r},\omega)$, 
have been explicitly performed
\cite{AMap}. 
A complete description of the classical electromagnetic fields of a 
radiating electric dipole has been achieved 
outside and inside a multi-structured spherical particle.
Electromagnetic fields have been 
determined anywhere in the space, and the time-averaged 
angular distribution of the radiated power, the time-averaged 
total radiated power, 
radiative and nonradiative (due to Ohmic losses) decay rates, 
frequency shifts have been calculated. 
Our recursive solution only employs $2\times 2$ transfer matrices and their
ordered products and provides a fast and reliable 
algorithm which can easily be implemented numerically \cite{AMap}.

In the present article, the theory developed in Ref. \cite{AMap}
will be applied to ``nano-matryoshka" structures of 
Prodan et al \cite{PRH}, i.e., multilayered spheres
consisting of alternating silica and gold concentric spherical shells. 
Such complex spheres 
have also been known as nanoshells.
Current experimental colloidal techniques allow one to design a 
variety of multi-structured beads having a plurality of concentric 
shells with the
core radius from cca 1 nm till 1 $\mu$m and controlled shell thicknesses.
For instance, metal (Au, Ag, Pt) and dielectric (ZnS) beads can be coated 
in a controlled way by a silica 
shell \cite{MGM,ULM,HMa,AvBV,VMB}, and a dielectric 
(silica, Au${}_2$S) bead can be coated 
by gold or some other noble metal
\cite{PRH,OAW,OJW,JaH,GAvB,CHM}.
One can subsequently etched away silica core of a silica-core 
metal-shell bead and obtain 
a hollow metallic nanoshell. Either hollow metallic 
nanoshell or a dielectric-core 
metal-shell bead can be in turn coated in a controlled way by the 
second concentric
silica shell (a dielectric overcoat 
of either metallic shell or a metallic core prevents aggregation 
of the particles by reducing the Van der Waals forces between them) 
\cite{PRH,GAvB} and by a further metal shell,
thereby forming a ``nano-matryoshka" \cite{PRH}. 
Compared to a simple homogeneous sphere, 
such a complex multilayered spherical particles 
allows one a lot more freedom in
engineering of both elastic \cite{PRH,OAW,OJW,JaH,NeB} and 
inelastic light-scattering properties \cite{JWH}.
Many other examples of stratified spheres can also be found in nature.
As an example, water insoluble aerosols in atmosphere have a 
thin liquid layer adsorbed on their surface. 
With the addition of an appropriate surfactant, water droplets in 
different hydrophobic solvents (such as oil) form a stable nanometer sized 
structures (with the size depending on the water to surfactant molar ratio), 
often referred
to as ``reversed micelles",
consisting of a spherical water core coated with a surfactant monolayer
 \cite{LRG}. 
In the case of a biological cell, 
the appropriate model consists of concentric three-layered sphere, 
corresponding to nucleus, cytoplasm, and membrane \cite{CKN}. 
The case of a sphere having 
two coatings is also important for modeling hydrological particles coated 
with biological material and micro-encapsulated material.

\begin{figure}[tbp]
\begin{center}
\epsfig{file=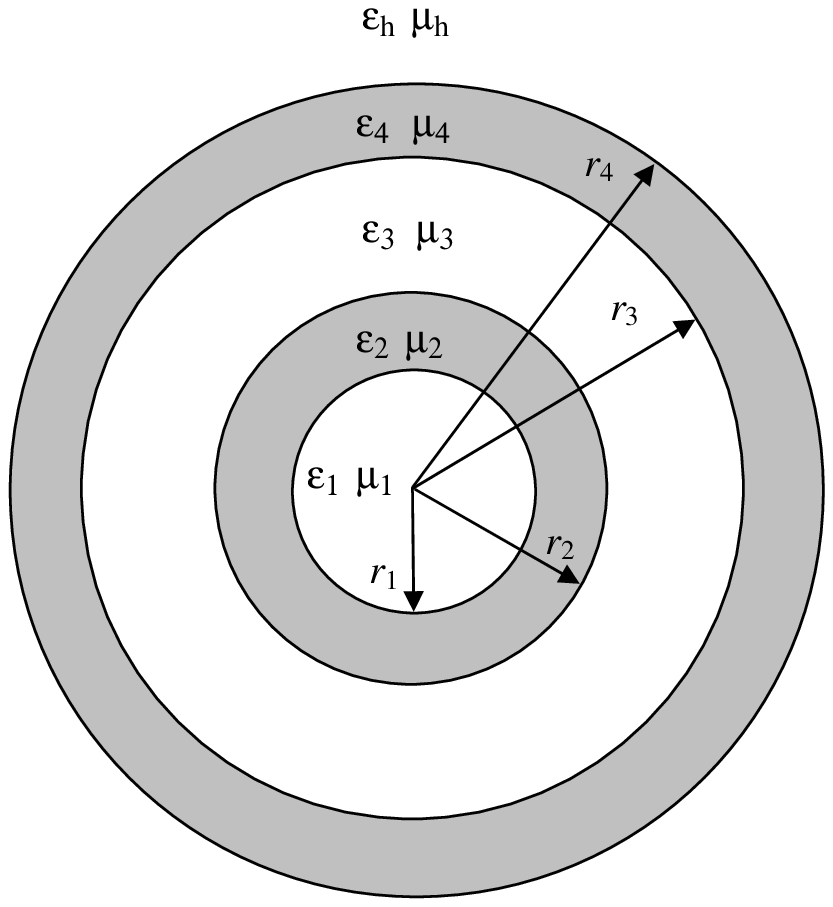,width=12cm,clip=0,angle=0}
\end{center}
\caption{}
\label{fgnaomif}
\end{figure}

A unique feature of ``nano-matryoshka" structures of 
Prodan et al \cite{PRH}, which partly explains
our focus on these structures, is the existence of two coupled nanocavities
surrounded by metal boundaries (see Fig. \ref{fgnaomif}).  
The latter feature holds promise of large and controlled
tunability of light-matter interactions, including both the ``nano-matryoshka" 
scattering properties and the spectroscopic properties of the atom
interacting with such a nanostructure. As it will be shown below,
such a complex nanoshell geometry strongly affects spontaneous 
emission decay rates, $W^{rad}$, photostability, and the
ratio $W^{rad}/W^{tot}$, known as the
fluorescence yield, or simply, quantum efficiency (here
 $W^{tot}=W^{rad}+W^{nrad}$ denotes the total decay rate) \cite{End}. 
The spontaneous emission rate $W^{rad}$ is one of three Einstein's coefficients
(usually denoted as A coefficient). The remaining two Einstein's coefficients
(usually denoted as B coefficients) describe
the {\em stimulated emission} probability 
and the {\em absorption} probability. When multiplied by the radiant
energy density $U_{\omega} d\omega$ with circular frequency 
between $\omega$ and $\omega +d\omega$, the respective 
B coefficients then determine the
stimulated emission and absorption rates at
 the transition frequency $\omega$.
Detailed balance in thermal equilibrium 
implies that the knowledge of a single Einstein coefficient is sufficient
to determine the remaining two. Therefore, once the spontaneous
emission rate is known, 
the stimulated emission and absorption rates
are also unambiguously determined.
For instance, the stimulated emission probability 
and the absorption probability are equal and the ratio of spontaneous 
to stimulated emission decay rate remains
equal to the mean thermal photon number at the transition frequency,
a constant which does not change with changing environment 
(see third reference in
\cite{Pu}). Thus, an inhibited (enhanced) spontaneous emission
necessarily implies inhibited (enhanced) stimulated 
emission by the same factor. 
There is hope that using complex nanostructures
one would be able to tailor spontaneous and stimulated decay rates 
according to one's need and a desired application, such
as chemical speciation, 
 LIDAR, fluorescent near-field microscopy, identification
of biological particles, and monitoring specific cell functions.
In this respect, fluorescent properties of the atom both inside
and outside of a complex nanostructure are of fundamental interest.
For instance, by placing fluorescent organic groups or rare earth 
ions (with nm control 
over the radial position \cite{AvBV}) inside 
the dielectric core or dielectric shell of 
such a complex nanoparticle \cite{CHM}, a fluorescent nanoprobe can
be formed for biophysical and biomedical applications \cite{Lak}.
Alternatively, for the applications in near-field optical microscopy 
\cite{BBN}, 
wherein
the probe tip is modeled as a sphere of small radius, one is interested 
in changes in the atom 
fluorescence properties induced by the presence of a nearby 
complex nanoparticle.

\section{Theory}
\label{sec:thr}
In order to characterize the change in the 
spectroscopic properties of the  atom interacting 
with a (complex) spherical scatterer, 
the frequency shifts, radiative and nonradiative decay rates 
will all be normalized with respect 
to the radiative decay rates of the same dipole but now 
in a free-space filled in with a homogeneous medium which
is identical to that at the atom position.
Such a normalization of spectroscopic properties brings an advantage
that, in the case of radiative decay rates, any local field corrections
\cite{RiK,LRG,Tom,SVL} cancel out.
In the case of a homogeneous dielectric medium (characterized by
the refractive index $n$ and dielectric permittivity $\varepsilon$)
Nienhuis and Alkemade \cite{NA} have showed that 
\begin{equation}
W^{rad}_h = \frac{n^3}{\varepsilon} W^{rad}_v,
\label{narel}
\end{equation}
where  $W^{rad}_h$ and  $W^{rad}_v$ are 
radiative rates of the electric-dipole transitions in the
dielectric medium and in the vacuum, respectively. 
For a non-magnetic medium, the above
relation reduces to $W^{rad} = n W^{rad}_0$.
It is emphasized here that these results holds
irrespective if a homogeneous medium is dispersive or not.
(For the general
case of a linear, nonconducting, absorptive, and dispersive medium see Sec. VIII
of Ref.
\cite{Tip}.)

Formally, irrespective of either the classical 
or quantum-mechanical descriptions, 
the line broadening and frequency shift of an electric-dipole
emitter interacting (via the vacuum electromagnetic field) with a material body
can be understood as a result of the coupling of the emitted field with its
own reflected field.
Let us label the concentric shells of a complex nanoshell
from the nanoshell core outward, 
with the nanoshell core counting as the shell number 
one and the ambient counting
as the shell number five.
Let $r_j$, $j=1,2,3,4$, and $\varepsilon_j$, $\mu_j$, 
$k_j=\omega_0 \sqrt{\varepsilon_j\mu_j}/c$, $j=1,2,3,4,5$,
denote the respective shell radii, dielectric permittivities, 
and wave vectors.
Occasionally, as in Fig. \ref{fgnaomif}, the subscript $5$ will 
be replaced by $h$ to indicate the host medium (ambient).
Then, within
the linear response 
formalism of Agarwal \cite{AgaIV} and of Wylie and Sipe \cite{WyS}, 
 the effective shift in the 
frequency separation $\omega_0$ of two levels of 
the  atom within the $n$-th shell
is given as follows
(see Refs. \cite{AgaIV,DKW,DKW1,CPS,KDLm}),  
\begin{equation}
\frac{\omega - \omega_0}{W_h^{rad}} = 
- \frac{3\varepsilon_n}{4 p^2 k^3_n}\, 
\mbox{Re}\, \left[ {\bf p}\cdot {\bf G} ({\bf r}_d,{\bf r}_d,\omega_0) 
\cdot {\bf p} \right]
= - \mbox{Re}\, \frac{3\varepsilon_n}{4p^2  k^3_n}\,{\bf p}
    \cdot {\bf E}_s ({\bf r}_d,\omega_0).
\label{qmshift}
\end{equation}
Here ${\bf r}_d$ is the dipole position, {\bf p} is the transition
dipole  moment,
$k_n=\omega_0 \sqrt{\varepsilon_n\mu_n}/c$, and
${\bf G}({\bf r},{\bf r}_d,\omega)$ denotes the {\em scattering} Greens function
normalized such that the electric field ${\bf E}_s({\bf r},\omega)$ 
of the scattered
radiation at ${\bf r}$ due to a dipole ${\bf p}$ radiating 
at frequency $\omega$ at ${\bf r}_d$ is given by
\begin{equation}
{\bf E}_s({\bf r},\omega) = {\bf G}({\bf r},{\bf r}_d,\omega)\cdot{\bf p}.
\label{grfnrm}
\end{equation}
For a homogeneous sphere one obtains in the 
{\em quasistatic limit} \cite{KDLm,KDL1}, 
\begin{eqnarray}
\left(\frac{\omega-\omega_0}{W_h^{rad}}\right)_\parallel &=&
 \frac{3}{32} \frac{\varepsilon_1 -\varepsilon_2} {\varepsilon_1 +\varepsilon_2} 
\frac{1}{ (k_2 r_s - k_d r_d)^3},
\nonumber\\
\left(\frac{\omega-\omega_0}{W_h^{rad}}\right)_\perp &=&
 \frac{3}{16} \frac{\varepsilon_1 -\varepsilon_2} {\varepsilon_1 +\varepsilon_2} 
\frac{1}{ (k_2 r_s - k_d r_d )^3},
\label{quasi}
\end{eqnarray}
where $k_d$ is the radiation wave vector in the medium wherefrom the dipole
is radiating.
The most characteristic feature of the quasi-static approximation
is that the frequency shift is a {\em monotonic} function of the
dipole distance from the sphere boundary.
Moreover, 
for the limiting cases of an atomic dipole in a close proximity
to the sphere and in the long-wavelength limit the quasi-static approximation
fails to account for the retardation effects and 
the radiative decay rate (linewidth) remains unchanged 
\cite{KDLm,KDL1}.

Whereas the change in the effective shift in the 
frequency separation of two levels
is given in terms of the {\em real} part 
of ${\bf G} ({\bf r},{\bf r},\omega_0)$, 
 the total decay rate induced by the 
presence of a (multilayered) sphere
is determined by the {\em imaginary} part of 
${\bf G}({\bf r},{\bf r},\omega_0)$.
Indeed, within
the linear response 
formalism of Agarwal \cite{AgaIV} and of Wylie and Sipe \cite{WyS}, 
the normalized decay rate for the  atom within the $n$-th shell
is given as \cite{AgaIV,DKW,DKW1,CPS,Ch,KDLm,Tom}
\begin{equation}
\frac{W^t}{W_h} = 1 +\frac{3\varepsilon_n}{2p^2 k^3_n}\, 
\mbox{Im}\, \left[ 
{\bf p}\cdot  {\bf G} ({\bf r}_d,{\bf r}_d,\omega_0)\cdot {\bf p} \right]
= 1 + \mbox{Im}\, \frac{3\varepsilon_n}{2 p^2 k^3_n}\, 
{\bf p}\cdot {\bf E}_s ({\bf r}_d,\omega_0).
\label{qmtrrate}
\end{equation}
The basic assumption is, of course, that neither the transition matrix 
element nor the transition frequency are appreciably changed by the presence 
of the interface.

In the presence of an absorption, as in our case, 
the decay rate $W^t$ 
 comprises the following two basic decay channels:
1) the process of real (i.e., not virtual) photon 
   emission with the photon escaping to the spatial infinity,
i.e., {\em radiative decay}, 
and
2) the process of real (i.e., not virtual) 
   photon emission accompanied by the 
   subsequent photon absorption by the microsphere, i.e., 
a {\em nonradiative decay}, 
\cite{AgaIV,DKW,DKW1,CPS,Ch,KDLm,Tom}.
In an ideal theoretical situation (i.e.,
a single fluorescent atom, the respective silica and gold shell
being without any impurities, and multi-photon relaxation absent)
the decay rate $W^t$, as calculated according to Eq. (\ref{qmtrrate}),
would be the total spontaneous decay rate. 
However,
the Ohmic loss is only one of many
other nonradiative mechanisms, such as, for instance, 
multiphoton relaxation, coupling
to defects, direct electron-transfer processes, and concentration 
quenching, which may 
all contribute to the nonradiative decay rate $W^{nrad}$ 
\cite{CEL,Bish,ITB,DSM,Dulk} but 
are not included in Eq. (\ref{qmtrrate}). 
Therefore, in practice, 
$W^t$ would be the lower limit to the total spontaneous 
decay rate $W^{tot}$. 

Obviously, in the absorbing case, 
the spontaneous decay rate $W^t$, 
as calculated according to Eq. (\ref{qmtrrate})
from the imaginary part of Green's 
function at coinciding arguments, ${\bf G}({\bf r},{\bf r},\omega)$, 
 \cite{DKW,DKW1}
does not coincide with the
radiative decay rate $W^{rad}$. 
The latter can  be, up to a proportionality factor, 
determined as the classically 
radiated power of a dipole which escapes to spatial infinity, 
or simply the radiative loss, $P^{rad}$,
which a classical dipole experiences when interacting 
with a (multicoated) sphere \cite{Ch,Chew,AMap,Rup}. 
$P^{rad}$ is calculated from 
the electromagnetic flux 
given by the surface integral of the Poynting vector 
through a virtual sphere of radius 
$R$ extending to infinity
\cite{Ch,Chew,AMap}.
According to the correspondence principle,
the radiative decay rates $W^{rad}$ is then given by 
\begin{equation}
W^{rad} =\frac{P^{rad}}{\hbar\omega}\cdot
\label{qmcm}
\end{equation}
In the absence of absorption, $W^{t}$ and $W^{rad}$
coincide. 
In the presence of an absorption, 
the  ratio $W^{rad}/W^t$, also known as {\em fluorescence yield},
is always smaller than one. The relative difference between  
$W^{rad}$ and $W^t$ is especially pronounced
in the proximity of metal boundaries (see Fig. \ref{fgnaomavy} below).
The quantum theoretical 
expression for the power radiated by the spontaneous 
emission from an excited state in 
an electric (a magnetic) dipole transition is 
still obtained from the classical 
expression for the power radiated by an electric (a magnetic) dipole, by 
replacement of the dipole moment by the corresponding transition matrix element.
(An expression for the dipole source intensity detected 
by a point detector
has been provided by Dung et al \cite{DKW1} (see Eqs. (34-36) therein).)

Let $k_0$ be the vacuum wave vector and $\varepsilon''$ ($\mu''$) 
be the imaginary part of the dielectric function 
(magnetic permeability) at the observation point.
The Ohmic loss contribution 
to the nonradiative
decay rates, $P^{nrad}$, is calculated
according to formula
\begin{equation}
P^{nrad} =\int_a Q({\bf r}) \, d{\bf r},
\label{pnrad}
\end{equation}
where the volume 
integral extends over all the absorbing regions.
$Q$ is given 
as the {\em steady} (averaged) inflow of energy per 
unit time and unit volume from the external sources which maintain the field,
\begin{equation}
Q=\frac{c k_0}{8\pi} \left(\varepsilon'' |{\bf E}|^2 +\mu'' |{\bf H}|^2 \right).
\label{endis}
\end{equation}
Here the {\em averaging} is performed with respect to time and
assuming that the amplitude of a monochromatic
electromagnetic field is a constant.
The formula (\ref{endis}) 
for Ohmic loss density remains also valid in the 
regions of high absorption near resonance frequencies of the 
permittivity and permeability, even 
when the so-called Brillouin expression for the 
electromagnetic field energy density,
\begin{equation}
U=\frac{1}{8\pi} 
\left[ {\bf E}\cdot{\bf E}^*\frac{d[\omega\varepsilon(\omega)]}{d\omega}+
{\bf H}\cdot{\bf H}^* \frac{d[\omega\mu(\omega)]}{d\omega}\right],
\label{brendns}
\end{equation}
 is no longer valid
(see Appendix C of Ref. \cite{AMap}). 
For simplicity, we will assume that $\mu''\equiv 0$, i.e.,  
the Ohmic losses will be entirely determined by an integral
of the squared amplitude of the electric intensity (see Ref. \cite{AMap}
for calculational details).

\section{Results}
\label{sec:res}
In this section, detailed results of numerical simulations are shown
for ``nano-matryoshka" structures of Prodan et al \cite{PRH}, i.e.,
multilayered spheres with silica core and surrounded by three
additional concentric spherical shells: an inner gold shell, 
followed by a silica spacer layer, and terminated by an outer gold shell. 
As in Ref. \cite{PRH},  ``nano-matryoshka" structures have been 
considered with the following
dimensions: $r_1/r_2/r_3/r_4$: $80/107/135/157$ nm
({\bf A}), $77/102/141/145$ nm ({\bf B}), and
 $396/418/654/693$ nm ({\bf C}). As comparative examples,
the results are also presented for 
a homogeneous silica sphere with radius $r_s=150$ nm ({\bf D}),
a homogeneous gold sphere with radius $r_s=693$ nm ({\bf E}),
and a homogeneous gold sphere with radius $r_s=150$ nm ({\bf F}).
The radius of the sphere {\bf E} was chosen to coincide with that
of the sphere  {\bf C}, whereas the radius of the spheres   {\bf D}
and {\bf F} was selected to lie between that of the spheres {\bf A} and {\bf B}. 
The radiating wavelength was taken to be $595$ nm, implying gold refractive
index $n_{Au} \approx 0.248+ i2.986$ \cite{Hand}.
All the spheres are assumed to be suspended in an aqueous solution.
The respective refractive indices of silica and water
are assumed to be $n_{SiO_2}=1.45$ and $n_{H_20}=1.33$.
There is nothing particular in choosing the wavelength of $595$ nm,
except of it being an emission wavelength of lissamine molecules \cite{Dulk}.
Any other (nonresonant) optical wavelength would lead to qualitatively 
similar conclusions. The choice of water as an ambient has been motivated
by a fact that (i) this often corresponds to an experimental situation
and (ii) aqueous solution matches biological conditions.

In numerical simulations, the angular-momentum cutoff value 
of $l_{max}=60$ was used.
In the case of the total and radiative decay rates, 
the cutoff value was sufficient to obtain convergence
on at least 8 significant digits (see Fig. 8 of Ref. \cite{AMap}). 
In the case of the Ohmic loss contribution, an immediate metal shell
proximity provides a numerical challenge. However further away from metal
interfaces the convergence of up to at least 8 significant digits can
be attained again (see Fig. 9 of Ref. \cite{AMap}).

\subsection{Frequency shifts}
\label{sec:shifts}
The  radiative frequency shifts have been calculated directly 
from the real part of Green's 
function at coinciding arguments according to Eq.
(\ref{qmshift}) (see also Eq. (137) of Ref. \cite{AMap}). The 
dependence of the frequency shifts
on an atomic dipole position
inside and outside the ``nano-matryoshka" structures {\bf A}-{\bf C}
and the homogeneous spheres {\bf D}-{\bf F} for the respective
 radial and tangential dipole orientations is shown 
in Figs. \ref{fgshperp}, \ref{fgshpar}.
Since in the sphere center the difference between the radial and tangential 
orientation of a dipole disappears, the corresponding radial and tangential 
quantities coincide there. 

In order to appreciate changes in the frequency shifts
induced by nanoshells {\bf A}-{\bf C} geometry, let us first discuss
the comparative example {\bf D} of a homogeneous silica sphere. 
{\em Inside} the dielectric sphere {\bf D},
the frequency shift of a radially oriented dipole steadily increases
from its value of $\approx 0.0117$ at the sphere origin till
that of $\approx 261$ at $r/r_s =0.995075$ 
(the last sampled point inside the spheres).
On the other hand,  
the relationship between the frequency 
shift and the position of a tangentially oriented
atomic dipole has an oscillating character,
first decreasing from the value of $\approx 0.0117$ at the sphere origin
down to $\approx 0.00784$ at $r/r_s =0.388$, and subsequently 
steadily increasing till $\approx 133$ at $r/r_s =0.995075$.
(Though due to the scale of ordinate axis, the shift
appears to be a flat featureless horizontal line).
{\em Outside} the dielectric sphere {\bf D}, it is
the frequency shift of the radially oriented dipole which
exhibits an oscillating behaviour, increasing
from the value of $\approx -334$ at the very first sampled point
outside the sphere at $r/r_s = 1.005025$, reaching
the maximum of  $\approx  0.0038$ at $r/r_s = 1.751294$ 
and then decreasing down to  $\approx  0.0021$ at the last sampled point at 
$r/r_s = 2.01$. For a tangentially oriented atomic dipole, the frequency shift 
gradually increases from the value of $\approx -162$ at $r/r_s = 1.005025$
up to  $\approx  -0.00017$ at $r/r_s = 2.01$ and compares
well with the quasi-static
approximation [see Eq. (\ref{quasi})]. 
The latter predicts a monotonic increase of frequency shifts
from large negative values to zero as $r/r_s$ increases.
Had the sphere radius was larger compared to the emission wavelength,
one would observe an oscillating relationship between the frequency 
shift and the position of a radiating
atomic dipole for all dipole orientations and both inside 
and outside the sphere. 
In agreement with the quasi-static result (\ref{quasi})
for a homogeneous dielectric sphere  
which is optically denser than surrounding
medium ($\varepsilon_s>\varepsilon_h$),
the frequency shift of an atomic dipole 
inside the sphere and in close proximity to 
its boundary is always toward {\em higher} frequencies 
({\em blue} shift) \cite{KDLm}. 
On the other hand, the frequency of an atomic dipole 
outside the sphere and in close proximity to 
its boundary experiences a shift toward {\em lower} frequencies 
({\em red} shift) \cite{KDL1}.

It is clear from Figs. \ref{fgshperp}, \ref{fgshpar} that  frequency shifts 
of an atomic dipole interacting with 
the ``nano-matryoshka" structures {\bf A}-{\bf C}
experience significantly enhanced changes than in the comparative
example of the homogeneous dielectric sphere {\bf D}. 
Already at the sphere center they
can be more than two orders of magnitude larger:
$\approx  -1.941$ for {\bf A}, $\approx -1.691$ for {\bf B}, 
and $\approx  0.903$ for {\bf C} compared to 
$\approx 0.0117$ for {\bf D}.
The emission frequency of an atomic dipole in a silica region inside 
the ``nano-matryoshka" structures {\bf A} and {\bf B}
is always shifted  toward {\em lower} frequencies 
({\em red} shift). 
For ``nano-matryoshka" structure {\bf C}, which radius 
is more than four times
larger than that of  {\bf A} and {\bf B}, 
the red shift is still observed in a proximity of gold shells. However,
in a marked contrast to ``nano-matryoshka" structures {\bf A} and {\bf B},
further away from gold shells, a small frequency shift 
toward {\em higher} frequencies 
is observed: for $r/r_s\in [0,0.17),\, (0.69,0.79)$  
and $r/r_s\in [0,0.23),\, (0.7,0.85)$ in the case of 
tangentially and  radially oriented dipole, respectively.
The magnitude of frequency shifts substantially
depends on the atom position
within a dielectric shell. 
With the atom approaching metal shell boundaries,
the frequencies exhibit an accelerated 
decrease toward large negative values.
We have seen
that even in  the case of a purely homogeneous dielectric
microsphere {\bf D} with a small refractive index contrast, 
the frequency shifts
are capable of reaching very high values near the surface of 
the microsphere.
However, for the respective ``nano-matryoshka" structures {\bf A}-{\bf B},
the shifts at the proximity of metal-dielectric interfaces
can be more than two orders of magnitude larger.

Outside and
at a very close proximity of the outer sphere boundary,
large  {\em red} frequency shifts are observed for the complex nanoshells
{\bf A}-{\bf C} as well as for the homogeneous spheres {\bf D}-{\bf F}. 
For instance, for radially oriented dipole source at
 the very first sampled point
outside the spheres at $r/r_s = 1.005025$, 
this red shift 
ranges from $-7392$ ({\bf B}), through $-5858$ ({\bf F}),  $-5117$ ({\bf A}), 
$-334$ ({\bf D}), till $-67$ ({\bf C}, {\bf E}). For a
 tangential dipole orientation,  the red shift 
ranges from 
$-3597$ ({\bf B}), through $-2839$ ({\bf F}),  $-2480$ ({\bf A}), 
$-162$ ({\bf D}), till $-30$ ({\bf C}, {\bf E}). 
Note that for the largest ``nano-matryoshka" {\bf C},
and a homogeneous gold sphere {\bf F} of the same radius,
the comparable frequency shifts at the proximity of the
outer sphere boundary 
are $5$-times smaller than for the silica microsphere {\bf D}.
Surprisingly enough, the quasistatic result 
(\ref{quasi}) of Klimov et al \cite{KDLm,KDL1} that, 
in the close proximity
of a spherical shells interface,
the radial frequency shift  is approximately 
{\em twice} as large as the tangential frequency shift appears to 
apply also for complex nanoshells
(see Figs. \ref{fgshperp}, \ref{fgshpar}).

Further away from sphere boundaries, 
as the value of $r/r_s$ 
increases, the red shift typically changes into blue one and 
vice versa in dampened oscillations around zero.
According to Fig. \ref{fgshperp}, a noticeably large blue shift 
with maximum 
$\approx 2.14$ at $r/r_s \approx 1.154$
is observed for a radially oriented dipole source
{\em outside} ``nano-matryoshka"  {\bf B},
leading to {\em repulsive forces} between the atom and the dielectric 
microsphere \cite{KDL1}.
The blue shift persists in a large interval for $r/r_s \in (1.12, 1.78)$.
For a tangential dipole source orientation, this blue shift,
which occurs for $r/r_s \in (1.14, 1.275)$ and $(1.63, 2.01)$ with maximum 
$\approx 0.413$ at $r/r_s \approx 1.164$, becomes almost five-times
smaller.
Blue frequency shifts for the atom located outside a sphere are also
observed in the remaining cases, but they are almost one order of magnitude smaller.
For instance, for radially oriented dipole source,  
the  excursion above zero does not exceeds
$\approx 0.148$ for {\bf C} (at $r/r_s \approx 1.144$),
$\approx 0.146$ for {\bf E} (at $r/r_s \approx 1.154$),
$\approx 0.059$ for {\bf A} (at $r/r_s \approx 1.592$),
$\approx 0.033$ for {\bf F} (at $r/r_s \approx 1.741$),
$\approx 0.004$ for {\bf D} (at $r/r_s \approx 1.751$).
In the case of a tangentially oriented dipole source, 
the excursion above zero does not exceeds
$\approx 0.113$ for {\bf E} (at $r/r_s \approx 1.254$),
$\approx 0.104$ for  {\bf C} (at $r/r_s \approx 1.254$),
$\approx 0.062$ for  {\bf A} (at $r/r_s \approx 2.01$),
$\approx 0.05$ for  {\bf F} (at $r/r_s \approx 2.01$),
whereas for 
{\bf D} the frequency shift remains negative till $2.01$.
Hence, similarly to the case of the atom located either close to a plane or 
{\em inside} a homogeneous dielectric spheres  \cite{KDLm}, 
the relationship between the frequency 
shift and the position of the atomic dipole {\em outside} the spheres
considered here
has an oscillating character
The dampened oscillatory behaviour 
of the frequency shift for the radially oriented dipole  
outside the sphere {\bf D}
contradicts the conclusion reached 
by Klimov et al \cite{KDL1}. However, Klimov et al \cite{KDL1} have
only studied atoms at a distance from a dielectric sphere
not larger than $r/r_s \approx 1.2$, which is a too short distance to
observe any oscillating behaviour.
Had they drawn frequency shifts  for larger values of $r/r_s$, 
they might have observed the oscillatory behaviour, too.

\subsection{Decay rates, the Ohmic loss
contribution to nonradiative decay rates, and fluorescence yield}
\label{sec:rates}

\subsubsection{Total decay rate $W^t$}
The normalized  total decay rates $W^t$ as
calculated directly 
from the imaginary part of Green's 
function at coinciding arguments according to Eq. (\ref{qmtrrate}) 
(see also Eq. (135) of Ref. \cite{AMap}), are displayed
in Figs. \ref{fgperprm}, \ref{fgparrm}.
Obviously, one finds the rates in the sphere center identical
for the radial and tangential 
atomic dipole orientations.
An advantage in
dealing with the
normalized decay rates is that 
any local-field correction \cite{RiK,LRG,Tom,SVL} cancels out 
(see also Sec. \ref{sec:locc} below)
and, in principle, a
direct comparison between the normalized decay rates
 and experiment can be performed.

Similarly as in the preceding subsection,
in order to appreciate changes in the decay rates
induced by nanoshells {\bf A}-{\bf C} geometry, 
we will first discuss
the comparative example {\bf D} of a homogeneous silica sphere. 
{\em Inside} the dielectric sphere {\bf D},
the normalized  decay rate
for a radially oriented atomic dipole steadily decreases from
its maximum value of $\approx 0.94237$ at the sphere origin
down to  $\approx 0.83437$ at the last sampled point inside the sphere
at $r/r_s =0.995075$. 
On the other hand, the normalized  decay rate
for a tangentially oriented atomic dipole exhibits a weakly 
oscillating behaviour: it first increases from
the value of $\approx 0.94237$ at the sphere origin and reaches
its maximum of $\approx 0.95173$ at $r/r_s =0.497562$, and then 
 decreases 
down to  $\approx 0.90179$ at 
$r/r_s =0.995075$. 
Outside the dielectric sphere {\bf D}, the normalized  decay rate
for a radially oriented atomic dipole steadily decreases from
its maximum value of $\approx 1.27798$ at the first
 sampled point outside the sphere at 
$r/r_s =1.005025$ 
down to  $\approx 0.99896$ at the last sampled point at 
$r/r_s =2.01$. 
On the other hand, the normalized  decay rate
for a tangentially oriented atomic dipole exhibits a weakly 
oscillating behaviour: it first increases from
the value of $\approx 0.9824$ at $r/r_s =1.005025$, reaches
its maximum of $\approx 1.00414$ at $r/r_s =1.860746$, and then 
it decreases 
down to  $\approx 1.00366$ 
at the last sampled point outside the sphere at 
$r/r_s =2.01$. (Though due to the scale of ordinate axis, 
the decay rates
appear to be a flat featureless horizontal line in 
Figs. \ref{fgperprm}, \ref{fgparrm}).
Note in passing that for spheres of larger radius compared to the emission
wavelength one would observe an oscillating behaviour 
of decay rates for any atomic dipole orientation \cite{Ch,Chew}.

A characteristic feature of nanoshells
is a huge increase of the
decay rates for a dipole source in a close proximity to
metal boundaries, and especially when dipole is within the silica core.
In a purely dielectric case, such large decay rates are only observed
in the proximity of sharp resonances of large spheres \cite{Ch},
whereas in the present case they can be achieved with small nanospheres
without any special tuning to their internal resonances.
For the ``nano-matryoshka" structures {\bf A} and {\bf B}, the 
decay rates monotonically increase from the values at their
center of  $\approx 0.8751$
and $\approx 1.7979$, respectively, up to the respective
values at the last sampled core points of 
$\approx 2773$  ($5412$) and 
$\approx 2445$   ($4765$)  for the tangential (radial) dipole orientation.
The normalized  decay rates of an atomic dipole
at the core region of the largest ``nano-matryoshka" {\bf C}
exhibit a qualitatively different behaviour
which is characterized by pronounced minima 
$\approx 0.1696$ at $r/r_s =0.218955$    
and $\approx 0.1129$  at $r/r_s =0.209005$  
for the radial and tangential dipole orientations, respectively.  
Depending
on the nanoshell parameters and the atom position, 
both inhibited and enhanced  decay rates are observed, 
with the  decay rates maximum values (in hot spots) 
being between two and three orders of magnitude larger 
than the  decay rates minimum values (in cold spots) within the same shell. 
The positional sensitivity
of the  decay rates appears to be more pronounced
in the nanoshell core regions than in the second silica shell.

Outside the complex nanoshell {\bf C}, an oscillatory dependence of the
 decay rates on the dipole position is clearly visible.
The behaviour is closely matched by the case {\bf E}, i.e. the case of 
homogeneous metal sphere of the same radius.
The amplitude of the oscillatory dependence 
is much stronger than that discussed 
earlier for {\bf D}.
A pronounced oscillatory dependence of the
 decay rates on the dipole position outside the sphere
can also be seen for {\bf F} and a tangential dipole orientation.
For {\bf A} and a tangential oriented dipole interacting
with {\bf B} only a very weak oscillatory dependence is seen.
For  {\bf B} and a tangential dipole orientation 
and {\bf F} and the radial dipole orientation only monotonic
decrease of  decay rates is observed down to 
$\approx 0.10373$ and
$\approx 0.10188$, respectively,  at $r/r_s =2.01$

\subsubsection{Radiative decay rate $W^{rad}$}
The normalized {\em radiative} decay rates $W^{rad}$ as
calculated according to Eq. (\ref{qmcm}) are shown
in Figs. \ref{fgphotons1}
and \ref{fgphotons2}.
The radiative decay rate is more pronounced 
for a radially oriented atomic dipole, in which case
an order of magnitude enhancement can be expected for an optimal
atom position inside the second silica shell of the nanoshells 
{\bf A} and {\bf B}.
Outside the spheres, 
the largest (an order of magnitude) enhancement of 
the radiative decay rate 
is achieved in a proximity of the small homogeneous gold microsphere
{\bf F}, closely followed by the nanoshell 
{\bf A}. 
The best location of a tangentially oriented atomic dipole 
appears to be within the nanoshell {\bf B} silica core.
Whereas the radially oriented atomic dipole shows typically an enhanced
radiative decay rate outside the spheres, in the case  of
its tangentially orientation the radiative decay rate is generally
reduced. A strongly reduced radiative decay rate is also observed
for  an arbitrarily orientated atom  inside the silica shells
of the nano-matryoshka structure
{\bf C}. 
Further away from the spheres outer surfaces
the radiative decay rate shows dampened oscillations around one 
for any atomic dipole orientation.
The radiative decay rate  for a tangential dipole orientation 
exhibits a complex
behaviour within the nanoshell {\bf C} silica core, showing there
a local minimum of $\approx 0.0204$ at $r/r_s =0.199055$.

\subsubsection{Ohmic loss contribution to the decay rate}
Compared to the normalized total and radiative decay rates 
in Figs. \ref{fgperprm}, \ref{fgparrm}, \ref{fgphotons1},
and  \ref{fgphotons2},
the normalized Ohmic loss contribution to the decay rates shown
in Figs.
\ref{fgnrmperp}, \ref{fgnrmpar} does not exhibit any oscillatory 
behaviour outside the spheres. 
It is reminded here that the  Ohmic loss contribution to the  
decay rates has been calculated
according to Eqs. (\ref{pnrad}), (\ref{endis}). (Further computational details 
can be found in Secs. 6 and 8.2 of Ref. \cite{AMap}.)
In agreement with 
expectations, the Ohmic loss contribution 
steadily decreases down to zero with the increased atomic dipole
distance from the sphere surface.
It is also easily understandable that 
 the Ohmic loss contribution to the
decay rates rapidly increases when
the atom approaches metal boundary. 
However,
rather contra-intuitively, the Ohmic loss contribution 
within the nanoshell {\bf C} silica core decreases
from its maximum 
value of $\approx 0.2102$ at the center 
down to $\approx 0.044$ and $\approx 0.047$
for the  radial and tangential dipole orientations, respectively,
at the very last sampled silica core point at $r/r_s =0.567214$.     
For the radial dipole orientation, the Ohmic loss contribution 
 decreases monotonically, whereas for a tangential dipole orientation
it exhibits an oscillatory dependence with a series
of local minima and maxima:
first decreasing from the central value of 
$\approx 0.2102$ down to $\approx 0.086$ at $r/r_s =0.228905$, increasing
 up to $\approx 0.137$   at $r/r_s =0.398060$,  and eventually decreasing
down to  $\approx 0.047$ at $r/r_s =0.567214$.     
This effect appears to be real and not an artifact
of computational inaccuracies or numerical instabilities. 
Calculations in extended
precision yielded essentially the same result.
Note that the local minima of the decay rates and
 the Ohmic loss contribution 
to the  decay rate 
for a tangential dipole orientation occur
at more or less the same position within the nanoshell {\bf C} silica core. 

\subsubsection{Fluorescence  yield}
In Fig. \ref{fgnaomavy} the ratio 
$W^{rad}/W^{t}$ is plotted, which
is known as the
{\em fluorescence  yield}, or simply {\em quantum efficiency}.
Our $W^{t}$ is that calculated directly 
from the imaginary part of Green's 
function at coinciding arguments according to Eq. (\ref{qmtrrate})
and has earlier been shown in Figs. \ref{fgperprm}, \ref{fgparrm}.
However, our 
$W^{t}$ only comprises the Ohmic loss contribution, $W^{nrad}_{\Omega}$, 
to the nonradiative decay rate $W^{nrad}$.
The Ohmic loss is only one of many nonradiative mechanisms, 
such as, for instance, 
multiphoton relaxation, coupling
to defects, direct electron-transfer processes, and concentration 
quenching, which
all contribute to the nonradiative decay rate $W^{nrad}$ 
\cite{CEL,Bish,ITB,DSM,Dulk} and which may all occur at real experimental
situations. 
Therefore, our 
$W^{t}=W^{rad} + W^{nrad}_{\Omega}$
does only provide the lower bound for the total decay rate $W^{tot}$.
Since 
$W^{rad}/W^{tot} \leq W^{rad}/ (W^{rad} + W^{nrad}_{\Omega})$,
the ratio $W^{rad}/W^{t}$ for an averaged 
 dipole orientation plotted in Fig. \ref{fgnaomavy} 
provides an upper theoretical bound 
on the fluorescence yield.  
Because of absorbing bodies, $W^{nrad}_{\Omega}>0$ and 
the respective ratios $W^{rad}/W^{t}$ are 
always smaller than one. (For the sake of clarity,
the trivial non-absorbing case {\bf D}, in which case the ratio is
equal to unity for any atom position, has been omitted.)
As expected,
with increasing the atom distance from the outer sphere surface,
where the Ohmic 
loss contribution $W^{nrad}_{\Omega}$ 
decreases to zero, the ratio $W^{rad}/W^{t}$ rapidly
approaches unity.
For all
cases considered, the respective 
fluorescence  yields are already larger than $0.93$ at $r/r_s=2$. 
At nanoshell centers, a remarkably large value of the fluorescence  yield
($\approx 0.694$) is observed for nanoshell {\bf B}. 
In the case of nanoshell {\bf C},
the fluorescence  yield is the smallest ($\approx 0.160$).
As dipole approaches metallic shells, the respective
fluorescence  yields rapidly
drop to
very small values. 
This is well-known as {\em fluorescence quenching} \cite{Lak}.
In spite of decay rates being strongly enhanced 
in the proximity of metal
shells (see Figs. \ref{fgperprm}, \ref{fgparrm}), 
the decay of the excited atomic states
is not accompanied by the emission of a real photon,
but instead matter quanta are created due to absorption.
The fluorescence quenching at the proximity of metal boundaries
then implies pronounced maxima of the respective
 fluorescence  yields
when the atom is located at the middle of the second 
silica shell of nano-matryoshka structures {\bf A}-{\bf C}.
Indeed, the shell is surrounded by metal
shells on its both sides.
A noticeable feature is also a rather complex 
behaviour of the fluorescence  yield
within the core of nanoshell {\bf C}. In the latter case, 
we have already seen a complex 
behaviour in the case of decay rates and 
the Ohmic loss contribution to the decay rates.
The complex behaviour in all these cases is the result of 
almost $5\times$ larger core radius compared
to the core radii of nanoshells {\bf A} and {\bf B}.

\subsection{Photobleaching and 
detected intensity enhancement}
If one assumes that photobleaching of a dye takes
place only while the dye is in its excited states,
a sufficiently large enhancement of the spontaneous emission rates 
can significantly lower the
probability of switching into nonfluorescent dark (triplet) 
states, thereby 
increasing stability against photobleaching \cite{CHM,IME}.
The latter means that a fluorescent  dye molecule can emit more 
photons before irreversible chemical reactions prevent the
molecule from any further emission.
Let us assume that photostability is inversely
proportional to the excited state lifetime, i.e.,
proportional to the total decay rate \cite{End}. 
Then the photostability of the atom 
interacting with a nanoshell compared to that in the 
free-space filled in with a homogeneous medium which
is identical to that at the atom position
increases by the factor of
$W^{tot}/W^{rad}_{h}$. On the other hand, there is only 
the probability of $W^{rad}/W^{tot}$ that a given
decay will end up with photon at the spatial infinity.
Given the respective two situations,
let $N$ and $N_0$ be 
average numbers of the photons emitted by 
the fluorescence source (and  
{\em detected} in the far-field) till photobleaching. 
Then the ratio $N/N_0$ can be obtained as 
$(W^{tot}/W^{rad}_{h}) \times (W^{rad}/W^{tot})=W^{rad}/W^{rad}_{h}$,
i.e., the ratio is equal to the normalized radiative decay rate.
The latter has been plotted as a function of the 
source position in Figs. \ref{fgphotons1} 
and \ref{fgphotons2}.

The results presented so far have not assumed any incident radiation.
An atomic dipole has  already been assumed to be in an excited state
and the changes in its decay properties have been monitored. 
It is informative to assess the changes in the detected fluorescence
intensity due to the presence of a metal nanoshell.
By taking into account that 
\begin{enumerate} 
\item the respective stimulated emission and absorption
probabilities undergo the same changes in the presence of material
boundaries as the rate of the spontaneous emission;
\item the stimulated emission and absorption
 rates are given as the product of the intensity, or squared field strength, 
with the respective probabilities; 
\item the spontaneous decay rate $W^t$ can be enhanced
by more than $3$ orders of magnitude (see Figs. \ref{fgperprm}, \ref{fgparrm}),
the radiative decay rate $W^{rad}$ 
an order of magnitude (see Figs. \ref{fgphotons1}, \ref{fgphotons2}),
and intensity by $4$ or more orders of magnitude; 
\end{enumerate} 
one finds that an elementary absorption-emission cycle
can be accelerated by $7$ or more orders of magnitude, resulting
in the enhancement by $5$ or more orders of magnitude
of the detected fluorescence
intensity.

\section{Discussion}
\label{sec:disc}

\subsection{Size-dependent corrections and nonlocal effects} 
The calculations presented so far have been performed assuming
the bulk values of gold dielectric constant.
If theory presented here is to be applied for
a multilayered spherical particle with
a small metallic core or a thin metallic shell with a radius 
or thickness $S\lesssim 20$ nm
(such as the outer shell of nano-matryoshka {\bf B}), 
two effects have to be additionally considered. 
First, the bulk dielectric function
is modified, since the electronic mean free 
path is then shorter than in the bulk \cite{KG}. Second, 
nonlocal effects come into play \cite{MHa,Rupn,Leu}. The first effect
can  easily be incorporated by replacing 
the bulk dielectric function $\varepsilon_B(\omega)$ with its 
size-dependent modification
\begin{equation}
\varepsilon(\omega) = \varepsilon_B(\omega)  
+\frac{\omega_p^2}{\omega^2 +i\omega\tau^{-1}_B} 
- \frac{\omega_p^2}{\omega^2 +i\omega\tau^{-1}}\cdot
\label{bulkm}
\end{equation}
Here $\omega_p$ is the bulk plasmon frequency, $\tau_B$ is the relaxation time 
in the bulk metal, $\tau^{-1} = \tau^{-1}_B + v_F S^{-1}$ 
is the inverse relaxation time (also called damping coefficient $\Gamma$) 
corrected for the finite size of the particle,
and $v_F$ is the Fermi velocity. More generally,
\begin{equation}
\tau^{-1} = \tau^{-1}_B + A v_F S^{-1},
\label{tauf}
\end{equation}
where $A$ is a parameter determined by the geometry. 
For simple Drude theory and 
isotropic scattering one usually takes $A=1$. 

On the other hand, {\em nonlocal effects}, i.e., 
when the Fourier transform of
the dielectric function depends in addition to $\omega$ also on ${\bf k}$, 
are associated with the resonant excitation of {\em longitudinal} 
bulk plasmon modes (either
propagating ones, with frequency {\em above}  
the plasma frequency $\omega_p$, or
evanescent ones, with frequency {\em below}  
the plasma frequency $\omega_p$)
\cite{MHa,Rupn,Leu}. 
The neglect of the nonlocal responses of the substrate 
is the main reason why 
a phenomenological treatment will generally break down
when the radiating atom is very close (within a few nanometers)
to the sphere surface 
\cite{Leu}. Indeed, the spectroscopic properties 
of the two-level atom interacting with a complex
spherical nanoshell have been investigated 
here within the framework 
of {\em macroscopic} Maxwell's equations. The latter lead to unphysical
results in the limit $r\rightarrow r_s$, i.e.,
 when the atom approaches sphere surface.
For instance, the quasi-static
approximation [see Eq. (\ref{quasi})] predicts that frequency shifts
increase as $(r-r_s)^{-3}$. 
The  non locality brings about a natural cutoff pole order: the nonlocal 
sphere does not polarize significantly at angular momenta higher than a certain 
cutoff value $l_c$. The latter is of
principal importance since it allows for a fully converged treatment of 
multipolar excitation effects \cite{FuC}. Another important 
feature of the nonlocal dielectric function is that
it introduces a natural cutoff angular momentum for 
the excitation of near-field modes
\cite{Leu,FuC}, and thereby a natural cutoff for the 
convergence of the nonradiative decay rates.

The nonlocal effects 
may  significantly influence both the radiative and nonradiative rates of 
very small nanospheres and nano-matryoshka structures with very thin shells.
For instance, in the case of a small homogeneous nanosphere with radius 
of $r_s\leq 5$ nm, the radial
and tangential dipolar decay rates for low emission frequencies 
($\omega\leq 0.5\, \omega_p$) can be reduced by up 
to $2$ orders of magnitude
with respect to the local results \cite{Leu}.
On the other hand, 
for emission frequencies $\omega\geq 0.5 \omega_p$, the decay rates can be
up to $2$ orders of magnitude larger than in the local case \cite{Leu}. 
The theory of Leung \cite{Leu} has  been extended 
to the case of complex nanoshells (see Sec. 11.1 of Ref. \cite{AMap})
and its application will be presented elsewhere.

\subsection{Radiative decay rates vs
the local density of states}
In the ideal case of nonabsorbing and nondispersive material medium
(which, however (except for the vacuum), does not exist),
the trace of the imaginary part of the Green's function at coinciding
spatial arguments, which 
enters Eq. (\ref{qmtrrate}), is related to the local density of states,
$ \rho(\omega,{\bf r})
=-(1/\pi) \mbox{Tr Im}\, {\bf G} ({\bf r}_d,{\bf r}_d,\omega_0)$, 
where Tr denotes trace
(${\bf G}$ is a tensor quantity). 
This relation has prompted claims that 
radiative decay rates are proportional 
to the local density of states (LDOS).  
However, even in the ideal nonabsorbing 
homogeneous case 
one finds 
this claim true only when 
(i) one performs orientational averaging 
over atomic dipole orientations and (ii) properly includes medium dependent
prefactors [see Eq. (\ref{qmtrrate})]. 
Once dispersion comes into play, radiative decay rates cease
to be proportional to the LDOS.
Indeed, Nienhuis and Alkemade \cite{NA} have showed that 
the LDOS in the homogeneous dispersive case is
\begin{equation}
\rho(\omega) = n^2 \frac{d(\omega n)}{d\omega} \rho^{(0)}(\omega) ,
\label{nafct}
\end{equation}
where $n(\omega)=\sqrt{\varepsilon\mu}$ is the frequency dependent 
refractive index
of the medium and $\rho^{(0)}(\omega)=\omega^2/ c^3\pi^2$ 
is the LDOS of photon states in vacuum (see Eqs. (29-31) of 
Ref. \cite{NA}). 
Yet, as it has been emphasized, the Nienhuis and Alkemade formula 
(\ref{narel}), which does not contain any derivatives
of the medium refractive index, remains also to be valid 
in the dispersive case. 

The presence of a derivative factor in the LDOS 
can also be traced down to similar derivative factors
in the Brillouin expression for the 
electromagnetic field energy density 
(\ref{brendns}). Another argument is as follows.
 Let $G(E)=1/(E-H)$ denote a scattering 
Green's function, where $E$ is an energy and
$H$ is a Hamiltonian.
Then, for an isolated eigenvalue $E_n$ of 
an energy dependent Hamiltonian (our case; see Refs. \cite{DKW,DKW1})
the quantity
$-(1/\pi)\mbox{Tr Im}\, G(E)$ is no 
longer equal to the Dirac delta function $\delta(E-E_n)$, 
but merely  proportional to $\delta(E-E_n)$ 
with a prefactor $1/|1 - dH/dE|$. The latter is, in general,
{\em different} from unity.
Therefore, the density of states (DOS)
$\rho(E)$ can no longer be defined as
$\rho(E) \equiv -(1/\pi)\mbox{Tr Im}\,  G(E)$.
Yet another argument is that the 
integrated density of states is given by the expression
$N(E) \equiv (1/\pi) \mbox{Tr Im}\, \ln G(E) 
= - (1/\pi) \mbox{Tr Im}\, \ln (E-H)$. 
Since, for an energy dependent Hamiltonian, the relation
$(d/dE) [ \mbox{Tr Im}\, \ln G(E)] =
 - \mbox{Tr Im}\,  G(E)$ no longer holds, 
$N(E) \neq \int^E \rho(E)\, dE$, provided
that the density of states (DOS)
$\rho(E)$ is defined as 
$\rho(E) \equiv -(1/\pi)\mbox{Tr Im}\,  G(E)$
 (see also Ref. \cite{AMip}).

\subsection{Local-field corrections}
\label{sec:locc}
An 
atomic dipole couples to the {\em microscopic} vacuum fluctuations. 
If the
(unnormalized) decay rate $W^{t}$ [see Eq. (\ref{qmtrrate})]
is
obtained from {\em macroscopic} Maxwell's equation, 
the difference between {\em microscopic} and {\em macroscopic}
vacuum fluctuations 
can be accounted for by {\em local-field corrections}.
The local field effects always play an important role 
and constitute 
a critical test of our 
understanding of the relation between microscopic and macroscopic 
electromagnetic phenomena \cite{RiK,LRG}.
As discussed by Schuurmans et al \cite{SVL}, one has
to distinguish 
between the substitutional and interstitial character of impurities.
In the former case  
the well-known empty-cavity factor $3\varepsilon/(2\varepsilon+1)$ applies, 
whereas, in the latter
case, the Lorentz local-field factor $(\varepsilon+2)/3$ is 
obtained. 
Since the decay rate $W^{t}$ can be expressed in terms of an 
expectation 
value of the
product of two electric field operators (see, e.g., Eq. (25) of Ref.
\cite{DKW1}), the local-field 
factors appear squared. 
In the near-infrared and for optical wavelengths, an 
inclusion of the corresponding local-field factor
is only necessary for electric dipole transitions.
Indeed, in the above wavelength range, 
magnetic permeabilities $\mu$ of different materials 
are all equal to the vacuum value and,
therefore, the local-field factors for magnetic dipole transitions
are trivially equal to one. (For local-field corrections 
in absorbing case see Ref. \cite{Tom}.)

However, since we have only discussed 
normalized decay rate $W^{t}$  [see Eq. (\ref{qmtrrate})], 
any local-field correction cancels out and a
direct comparison between the normalized decay rates
 and experiment can, in principle, be performed 
(see, however, the following
subsections).

\subsection{Nonradiative decay rates}
Earlier, at the end of Sec. \ref{sec:rates},
it has been discussed that 
 the sum $W^t=W^{rad} + W^{nrad}_{\Omega}$,
where $W^{nrad}_{\Omega}$ stands for 
the nonradiative decay rate due to the Ohmic losses,
does only provide the lower bound on the total decay rate $W^{tot}$.
Consequently, 
Fig. \ref{fgnaomavy} only shows 
the upper theoretical limits
on the fluorescence yields. Indeed,
the Ohmic loss is only one of many
other nonradiative mechanisms, such as, for instance, 
multiphoton relaxation, coupling
to defects, direct electron-transfer processes, and concentration 
quenching, which
all contribute to the nonradiative decay rate $W^{nrad}$ 
\cite{CEL,Bish,ITB,DSM,Dulk}. 
It turns out that even in a purely dielectric case, in the 
absence of any Ohmic losses and for small fluorescence atom
concentrations, 
the nonradiative decay $W^{nrad}$ can be higher than 
the radiative decay 
$W^{rad}$, resulting in the
fluorescence yield
smaller than $0.5$ \cite{LRG,DSM1,SLP}.
When fluorescence dye concentrations increase above a 
certain threshold value, 
the fluorescence yields of most organic dyes are
substantially reduced even further with respect to a 
zero-concentration limit value.
It should be emphasized
that, for a dipole outside the sphere,
the radiative decay rate $W^{rad}$ is proportional to the intensity
of the time-resolved fluorescence spectra at $t=0$ \cite{Dulk}.
Therefore, in the latter case, $W^{rad}$ and $W^{nrad}$ can be 
disentangled experimentally \cite{Dulk}.

In order to test theoretical predictions experimentally, the choice of
a suitable fluorescence source turns out to be a 
critical issue. Rare-earth ions, which exhibit long luminescence lifetimes,
 are very suitable candidates 
\cite{RiK,DSM,SSa}. However,
they are usually implanted by an ion deposition resulting in a poor
control over their radial distribution within a spherical shell.
On the other hand, 
fluorescent organic groups can be placed inside 
the dielectric core or shell of a complex nanoparticle with nanometer control 
over the radial position \cite{AvBV}, but their decay rates
are usually strongly affected also by other then purely electromagnetic
mechanisms, such as concentration quenching. 
For instance, in the case of fluorescein (FITC),
the threshold value for a concentration quenching 
is $\sim 0.1$ mM ($\sim $1 mM) in liquid (solid) solutions \cite{IME}.
However pyrene-doped PMMA spheres with pyrene concentrations up to 
$10$ mM do not exhibit any concentration dependence 
\cite{FSM}.
Note also that metals per se have inherent photoluminescence \cite{Mo}.
However, since
metal photoluminescence results in a broad band continuum \cite{BBN,Mo},
it can  easily be filtered out.

\subsection{Nonlocal decay rates} 
The nonradiative mechanisms different 
from the Ohmic losses do only depend on the immediate neighborhood 
of a radiating dipole. Therefore, one can
include all such mechanisms of nonradiative decay rates under the 
{\em local} nonradiative decay rate, $W^{loc}$. On the other hand,
the nonradiative decay rate due to Ohmic losses and the radiative decay rate
can be viewed as a
{\em nonlocal decay rate}, $W^{nloc}\equiv W^t=W^{rad}+ W^{nrad}_{\Omega}$. 
The reason behind this nomenclature is that the
latter two rates depend on the geometry and material
composition of the entire sphere and surrounding medium, and not 
only on the immediate proximity of the radiating dipole. 
The total decay rate is then written as
$W^{tot}= W^{loc} + W^{nloc}$.
As a rule,
both the nonradiative decay rate due to Ohmic losses 
and the radiative decay rate
change if the optical properties
of a shell being far away from the radiating dipole
(for instance, ambient medium) change \cite{DSM,DSM1,SLP}. 
In the case of a homogeneous dielectric sphere, the local and nonlocal 
decay rates can be, in principle, disentangled by measuring the total decay
rate $W^{tot}$ using the same sphere 
in different environments \cite{RiK,LRG,DSM1}. For instance, the sphere can be
embedded in a refractive index matched liquid \cite{RiK,LRG,DSM1}.
Its radiative decay rates then becomes that of a dipole in a homogeneous 
dielectric slab \cite{SLP,UR}. Other possible environments
include air or liquids with different refractive indices \cite{DSM1}.
The difference of the respective total decay rates measured in different sphere
environments then corresponds to the
difference of the nonlocal decay rates \cite{DSM1}. The latter is
obviously also true for a multicoated sphere.
The local and nonlocal decay rates can then be separated by fit of Eq. 
$W^{tot}= W^{loc} + W^{nloc}$
to the measured data \cite{DSM,DSM1}. Only after the 
{\em local} decay rate $W^{loc}$ is determined, 
a comparison of measured data and theory presented here can be performed
\cite{CHM}.

\subsection{Numerical subtleties} 
When dealing with 
dispersive and absorbing shells, it may happen that 
the linearly independent spherical Bessel functions 
$j_l$ and $y_l$ (see Sec. 10 of Ref. \cite{AS} for 
notations) are in fact related by $y_l(kr)\approx ij_l(kr)$
up to almost all significant digits in double precision. 
Consequently, if the spherical Hankel functions $h_l^{(1)}$ is 
later on formed as the sum $h_l^{(1)}(kr)=j_l (kr)+ iy_l(kr)$ \cite{AS},
its precision may be drastically compromised. 
Therefore, it is always recommended
to determine $h_l^{(1)}(kr)$ by a direct independent recurrence, 
such as that proposed by 
Mackowski et al \cite{MAM} [see recurrences (63),(64) therein].
Otherwise the radiative decay result for the interior of the nanoshell
{\bf C} may differ by up to {\em four-orders} of magnitude from the correct one.
If one can perform calculations in an extended precision, 
this pathological behaviour can be largely overcome, yet the independent 
recurrence by Mackowski et al \cite{MAM} is still highly recommended.

\subsection{Outlook} 
Interesting avenue of further research is the study of the effect of 
sphere's resonances on the atomic spectroscopic
properties. So far, the latter problem has been thoroughly investigated 
only in the
case of a homogeneous dielectric sphere \cite{KDL1}
and a dispersive and absorbing sphere \cite{DKW1}. 
However, a comparative study for complex nanoshells
is missing. Yet another interesting set of problems arises
in connection with strong atom-sphere interactions, leading,
for instance, to  Rabi splitting of an atom levels.
Also here only the case of a dielectric homogeneous 
microsphere has been considered \cite{KDL2}. 

Another important aspect
is the inclusion of nonlocal effects into the treatment of spectroscopic
properties of the atom interacting with complex nanoshells.
As it has been discussed earlier, 
the neglect of the nonlocal responses of the substrate 
is the main reason why 
a phenomenological treatment will generally break down
when the radiating atom is very close (within a few nanometers)
to the sphere surface \cite{Leu}.
So far, the effect has only been investigated in the case of small
metal nanospheres \cite{Leu}. In this regard,
complex "nano-matryoshka" structures offer the possibility
in studying the nonlocal effects
for spheres of relatively large radius, provided that metal
shells are thin enough. General theory for the case of "nonlocal"
shells has already been developed \cite{AMap} and its application
will be presented elsewhere.

\section{Summary and conclusions}
\label{sec:concl}
\begin{figure}[tbp]
\begin{center}
\epsfig{file=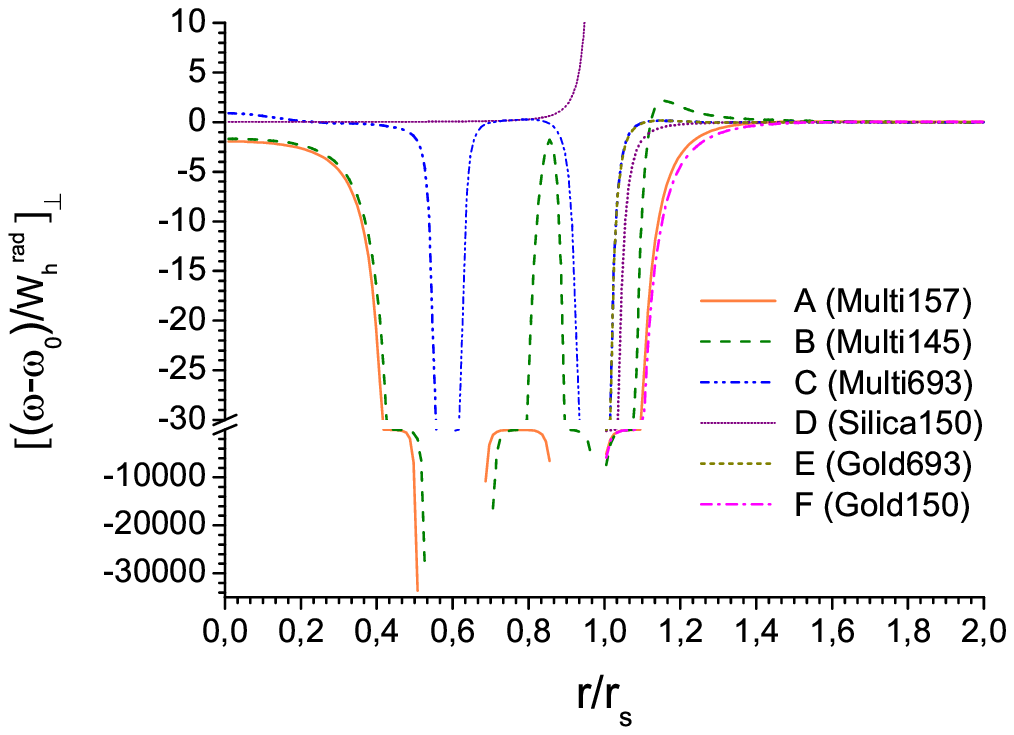,width=12cm,clip=0,angle=0}
\end{center}
\caption{}
\label{fgshperp}
\end{figure}

\begin{figure}[tbp]
\begin{center}
\epsfig{file=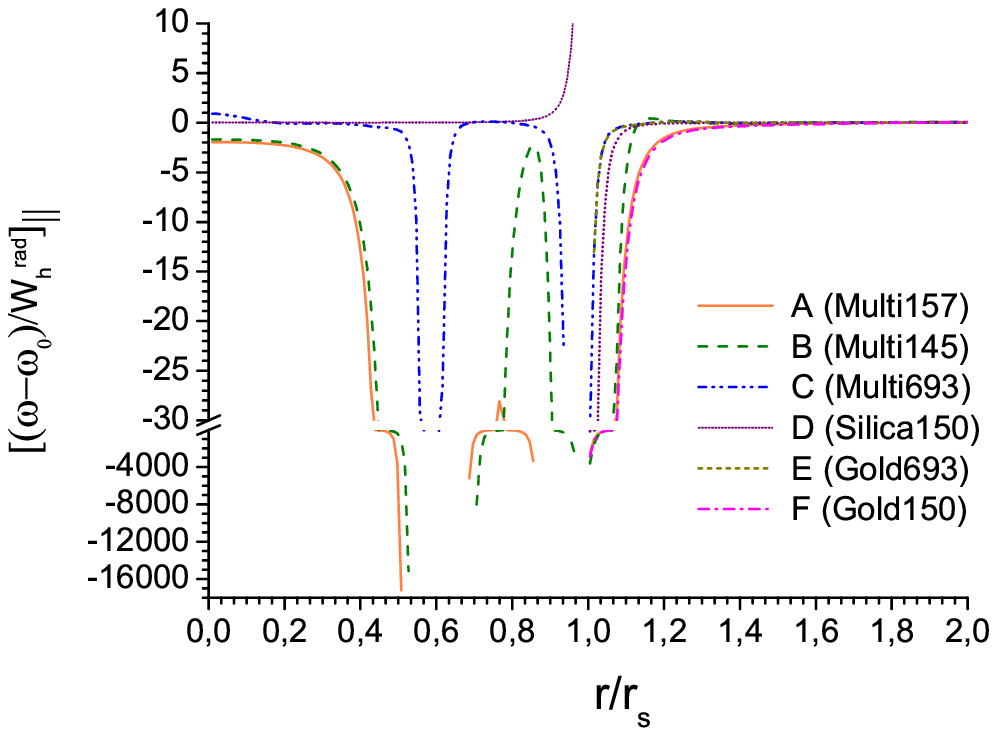,width=12cm,clip=0,angle=0}
\end{center}
\caption{}
\label{fgshpar}
\end{figure}

Frequency shifts, decay rates, 
the Ohmic loss contribution to the nonradiative decay rates,
fluorescence yields, and
photobleaching of 
an atomic dipole interacting with
``nano-matryoshka" structures of Prodan et al \cite{PRH} have been 
investigated. 
The changes in the spectroscopic properties of an atomic 
dipole interacting with
complex nanoshells have been shown to be significantly enhanced, often 
more than two orders of magnitude, compared
to the same atom interacting with a homogeneous 
dielectric sphere. 
The detected fluorescence intensity 
can be then enhanced by $5$ or more
orders of magnitude. The decay rate enhancements can be 
achieved with small nanospheres
without any special tuning to their internal resonances.
The changes strongly depend
on the nanoshell parameters and the atom position. 
Rather contra-intuitively the Ohmic loss contribution 
to the nonradiative decay rates
for an atomic dipole within the silica core 
of larger nanoshells may be decreasing when
the silica core - inner gold shell interface is approached. 
Surprisingly enough, the quasistatic result 
(\ref{quasi}) of Klimov et al \cite{KDLm,KDL1} that 
the radial frequency shift in the close proximity
of a spherical shell interface is approximately 
{\em twice} as large as the tangential frequency shift appears to also
apply for complex nanoshells
(see Figs. \ref{fgshperp}, \ref{fgshpar}).
Although decay rates are strongly enhanced in the proximity of metal
shells, the majority of the emitted radiation 
is absorbed and fluorescence yield exhibits there a well-known
quenching (see Fig. \ref{fgnaomavy}).
Although simulation have so far been performed at the wavelength of $595$ nm
(the emission wavelength of lissamine molecules \cite{Dulk}),
Fig. \ref{fgtotAld} demonstrates that
significantly modified spectroscopic properties can be observed in a broad band
comprising all (nonresonant)  optical
and near-infrared wavelengths.

It has been shown here that complex nanoshell structures 
provide the possibility of
a controlled tunability in 
engineering of the radiative decay rates.
The ability of controlled modification of radiative rates for atoms 
or molecules in the excited
state is of great importance since dissipative pathways of the excited state 
can be controlled. For instance, one can design
nanoprobes  with enhanced quantum yield for 
fluorescent microscopy and with enhanced photostability for 
for biophysical and 
biomedical applications \cite{Lak,CHM}, identification of biological particles in
fluorescence-activated flow of cytomeres \cite{MaGl},
to monitor specific cell functions, or in the cell
identification and sorting systems \cite{MuD,YaC}. 
Enhanced spontaneous emission rates 
could also provide increased sensitivity in low level
fluorescence applications \cite{Bish,IME}. Designing of 
small noble metal 
nanoparticles with reduced quantum yield, $W^{rad}/W^{tot}$, 
at the particle close proximity (cca $1$ nm)
may have crucial implications for 
the particles use as acceptors in biophysical F\"{o}rster
resonant energy transfer experiments in vitro as well as in vivo \cite{Dulk}.
The theory presented earlier in Ref. \cite{AMap} with numerical results
shown here may also stimulate  design of coated tip geometries for applications
in near-field optical microscopy \cite{BBN}.

\begin{figure}[tbp]
\begin{center}
\epsfig{file=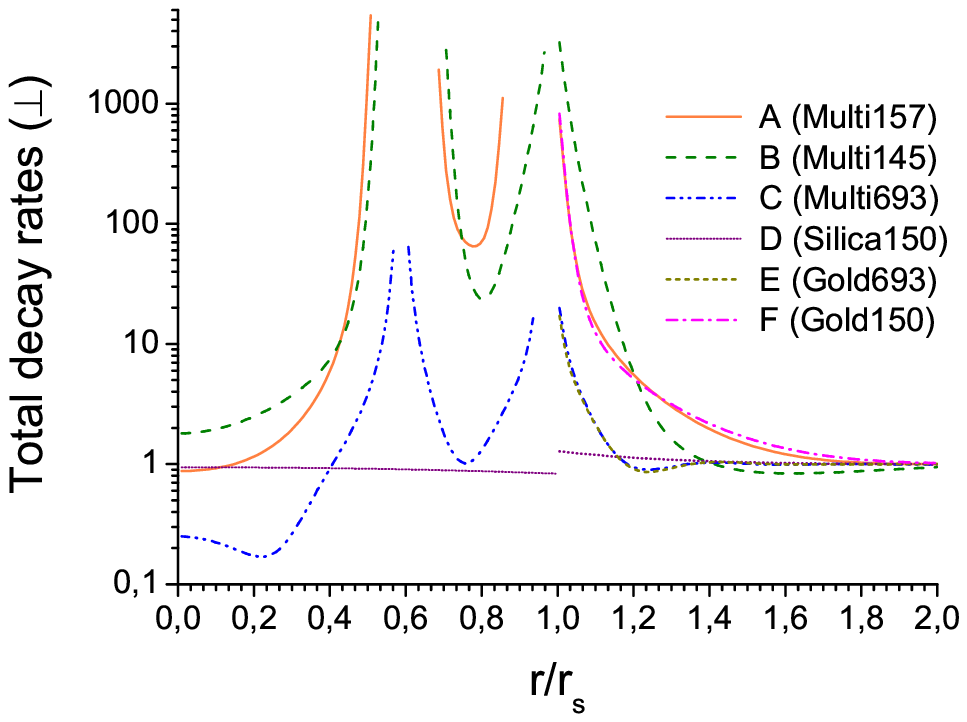,width=12cm,clip=0,angle=0}
\end{center}
\caption{}
\label{fgperprm}
\end{figure}

\begin{figure}[tbp]
\begin{center}
\epsfig{file=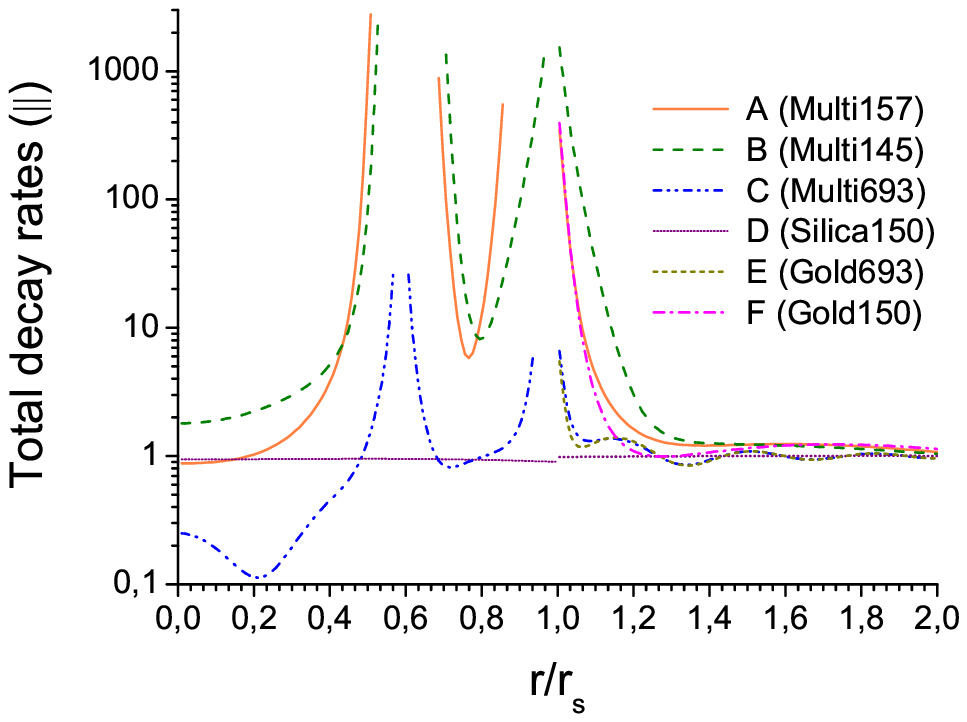,width=12cm,clip=0,angle=0}
\end{center}
\caption{}
\label{fgparrm}
\end{figure}

The emphasis was on the spontaneous decay, rates. However, 
detailed balance in thermal equilibrium 
implies that the knowledge of a single Einstein coefficient is sufficient
to determine the remaining two. Therefore, qualitatively similar behaviour 
as that shown in Figs. \ref{fgperprm}, \ref{fgparrm} is also expected for 
the {\em stimulated emission} and {\em absorption} rates.

Hopefully, the results presented in this article,
in conjunction with computer program freely available at 
http://www.wave-scattering.com,
will provide a larger freedom
in engineering of (complex) spherical particles properties, 
rendering them more suitable
for a variety of applications.

\newpage

\begin{center}
{\large\bf Figure captions}
\end{center}

\vspace*{2cm}

\noindent {\bf Figure 1 -} A typical spherical complex nanoshell,
or ``nano-matryoshka", and its parameters. In the present case, 
the ``nano-matryoshka" will be embedded in water and
the respective shaded and unshaded ``nano-matryoshka" concentric 
regions will represent gold and silica shells.

\noindent {\bf Figure 2 -} The normalized level shifts
for the radially oriented atomic dipole
radiating at wavelength of 595 nm as a function of its distance from
the sphere center. The shifts are normalized to the 
radiative decay rate $W_h^{rad}$ in free-space filled in with the 
medium at the dipole position. For the homogeneous dielectric sphere 
{\bf D}, the frequency shift reaches the value as large as 
$261$ at $r/r_s=0.995075$. However, for a better view of the region around
zero frequency shift, the ordinate axis has been terminated
at the frequency shift of $10$.

\noindent {\bf Figure 3 -} The same as in Fig. \ref{fgshperp} but for 
tangentially oriented atomic dipole. For the homogeneous dielectric sphere 
{\bf D}, the frequency shift reaches the value as large as 
$133$ at $r/r_s=0.995075$. However, as in Fig. \ref{fgshperp},
for a better view of the region around
zero frequency shift, the ordinate axis has been terminated
at the frequency shift of $10$.

\noindent {\bf Figure 4 -} Normalized decay rates 
$W^{t}/W^{rad}_h$ for the radially 
oriented  atomic dipole
radiating at wavelength of 595 nm as a function of its distance from
the sphere center. The rates 
are normalized to the radiative decay rate $W_h^{rad}$ 
of the same atomic dipole in the free-space filled in with the 
medium at the dipole position.
 
\noindent {\bf Figure 5 -} The same as in Fig. \ref{fgperprm} but for 
tangentially oriented atomic dipole.

\noindent {\bf Figure 6 -} The normalized
radiative decay rate of the radially 
oriented 
fluorescence dipole source at wavelength of 595 nm 
 The rate has been normalized 
to that in free-space filled in with the 
medium at the source position.

\begin{figure}[tbp]
\begin{center}
\epsfig{file=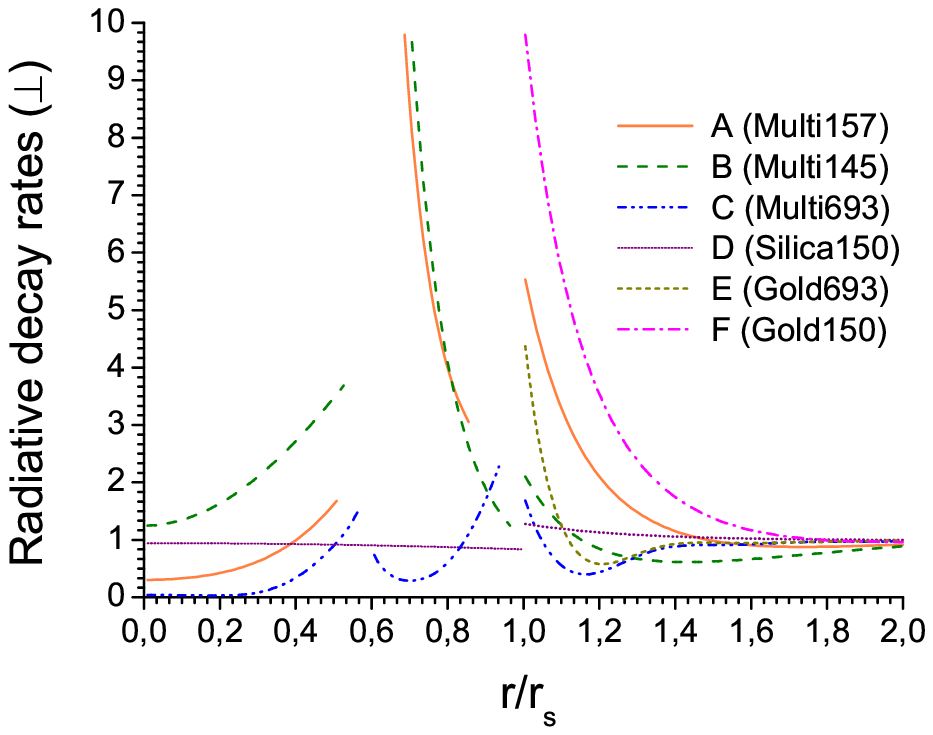,width=12cm,clip=0,angle=0}
\end{center}
\caption{}
\label{fgphotons1}
\end{figure}

\begin{figure}[tbp]
\begin{center}
\epsfig{file=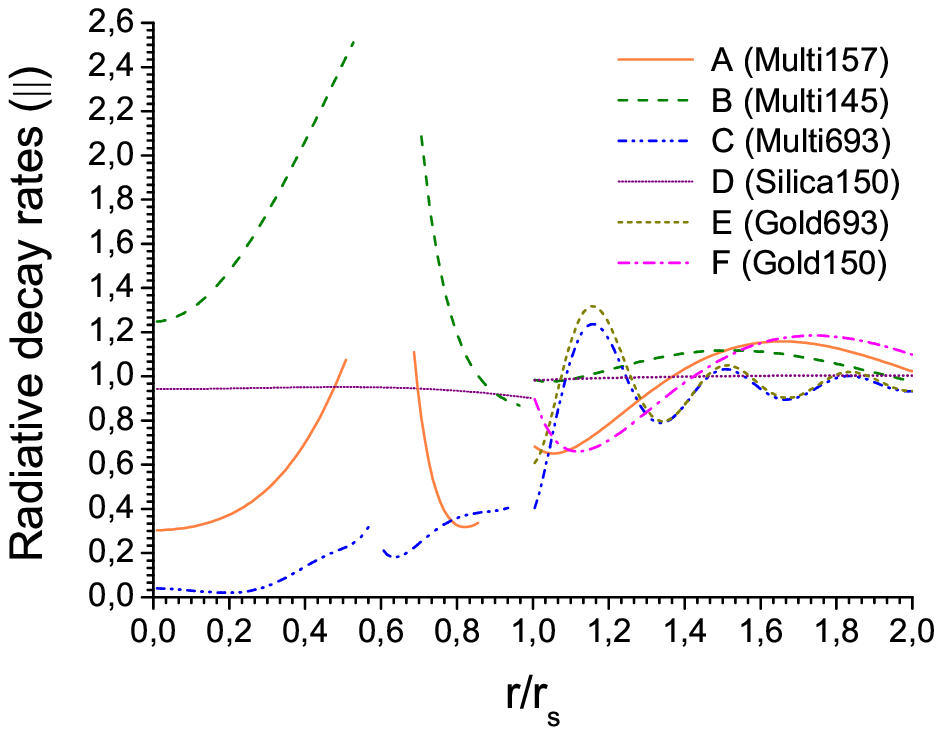,width=12cm,clip=0,angle=0}
\end{center}
\caption{}
\label{fgphotons2}
\end{figure}

\noindent {\bf Figure 7 -} The same as in Fig. \ref{fgphotons1}
but for 
tangentially oriented atomic dipole.

\noindent {\bf Figure 8 -} The normalized Ohmic loss contribution 
$W^{nrad}_{\Omega}/W^{rad}_h$
to the nonradiative
decay rates for radially oriented atomic dipole
radiating at wavelength of 595 nm as a function of its distance from
the sphere center. The Ohmic loss contribution has been normalized to the 
free-space radiative decay rate $W^{rad}_h$ in the 
medium identical to that at the dipole position.
 
\begin{figure}[tbp]
\begin{center}
\epsfig{file=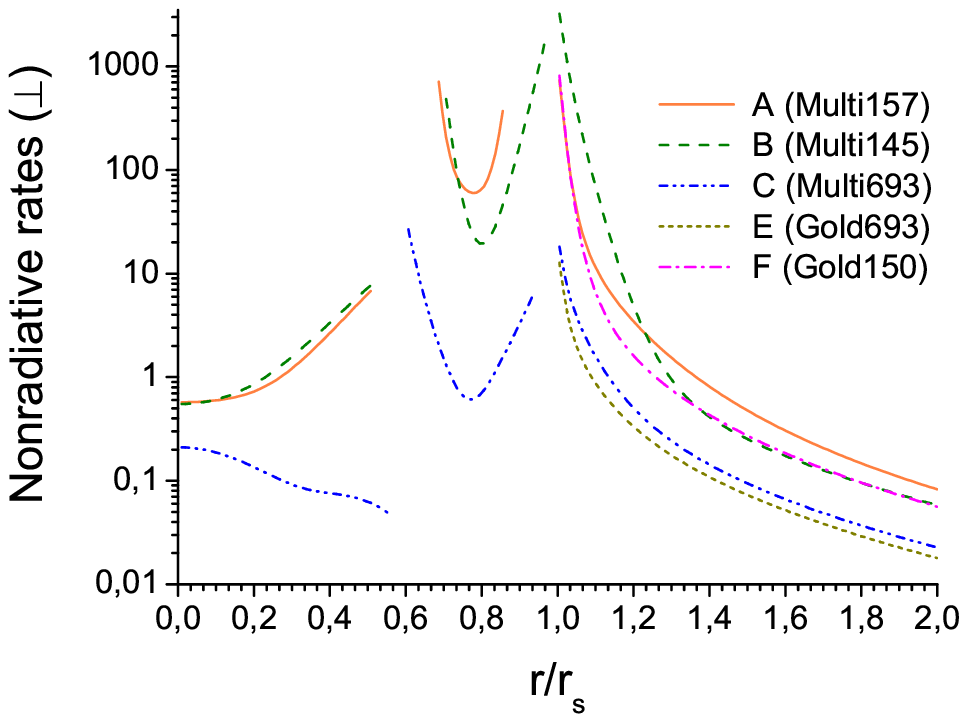,width=12cm,clip=0,angle=0}
\end{center}
\caption{}
\label{fgnrmperp}
\end{figure}

\begin{figure}[tbp]
\begin{center}
\epsfig{file=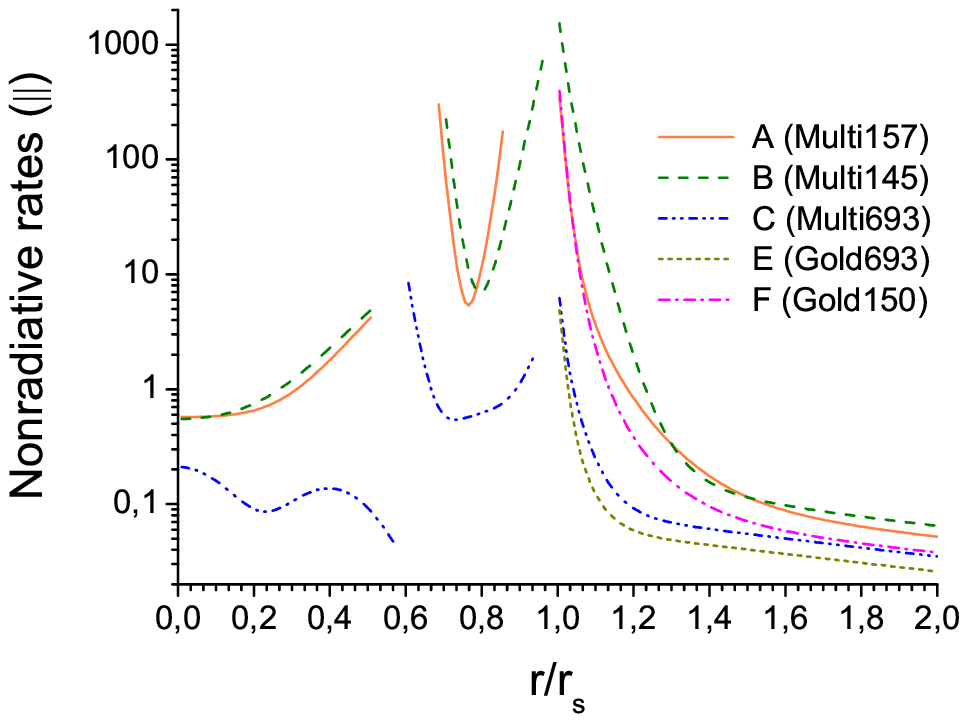,width=12cm,clip=0,angle=0}
\end{center}
\caption{}
\label{fgnrmpar}
\end{figure}

\begin{figure}[tbp]
\begin{center}
\epsfig{file=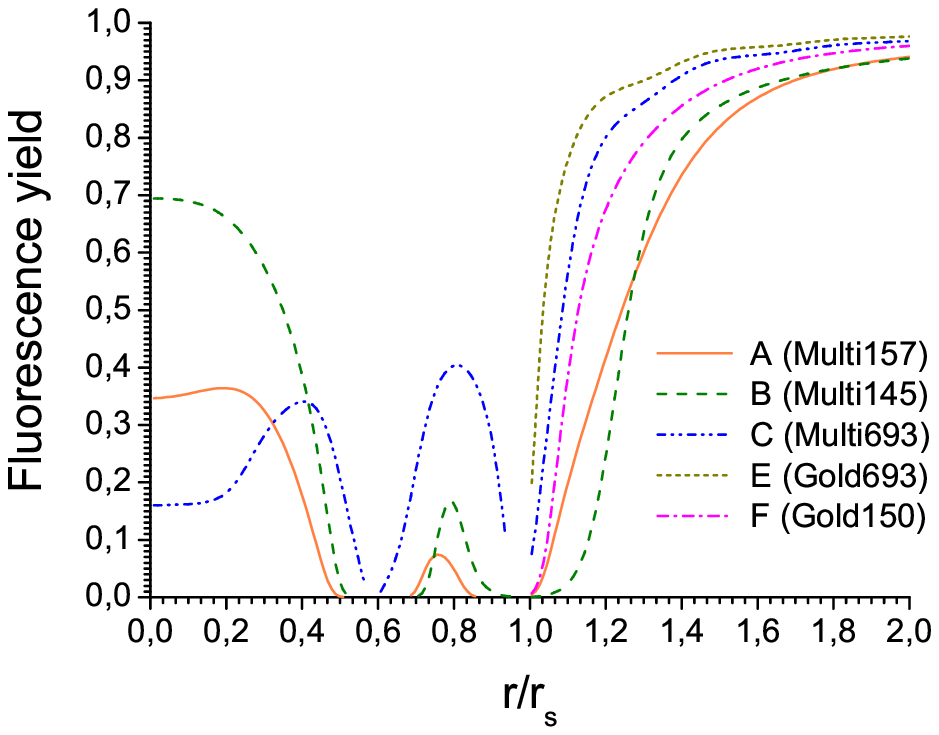,width=12cm,clip=0,angle=0}
\end{center}
\caption{}
\label{fgnaomavy}
\end{figure}

\begin{figure}[tbp]
\begin{center}
\epsfig{file=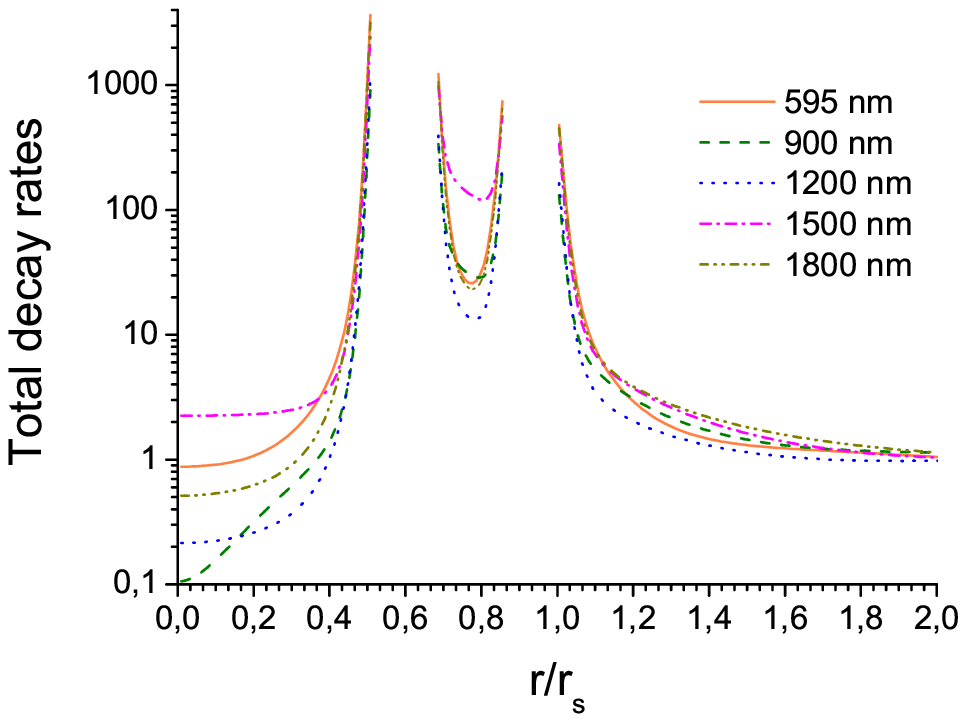,width=12cm,clip=0,angle=0}
\end{center}
\caption{}
\label{fgtotAld}
\end{figure}

\noindent {\bf Figure 9 -} The same as in Fig. \ref{fgnrmperp} but for 
tangentially oriented atomic dipole.

\noindent {\bf Figure 10 -} Fluorescence
yield (quantum efficiency) at wavelength of 595 nm for an averaged 
dipole orientation as a function of the 
dipole position.

\noindent {\bf Figure 11 -} Normalized decay rates 
$W^{t}/W^{rad}_h$ for an averaged 
dipole orientation in the case of nanoshell {\bf A} 
as a function of the dipole position for different emission
wavelengths.

\newpage

\newpage

\newpage

\newpage

\newpage

\newpage

\newpage

\newpage

\newpage

\newpage

\newpage

\end{document}